\RequirePackage[2020-02-02]{latexrelease}
\documentclass[prb,a4paper,aps,10pt,twocolumn,floatfix,superscriptaddress,amsfonts,longbibliography]{revtex4}
\pdfoutput=1
\usepackage[T1]{fontenc}\usepackage[latin1]{inputenc}
\usepackage{dcolumn,graphicx,color,booktabs,microtype,afterpage} 
\usepackage{sidecap}
\usepackage{times}
\usepackage[colorlinks,plainpages=false,linkcolor=blue,urlcolor=blue,citecolor=blue,pdfpagemode=UseNone,pdfstartview=FitBH]{hyperref}

\usepackage{amssymb}
\usepackage{amsmath}
\usepackage{physics}
\usepackage{graphicx}
\usepackage{esvect}
\usepackage{bm}
\usepackage{tabularx}
\newcommand{\ceth}{Ce$^{3+}$}
\newcommand{\cefo}{Ce$^{4+}$}

\newcommand{\snfo}{Sn$^{4+}$}
\newcommand{\cesno}{Ce$_2$Sn$_2$O$_7$}
\newcommand{\cezro}{Ce$_2$Zr$_2$O$_7$}

\newcommand{\ndzro}{Nd$_2$Zr$_2$O$_7$}
\newcommand{\ybtio}{Yb$_2$Ti$_2$O$_7$}
\newcommand{\hotio}{Ho$_2$Ti$_2$O$_7$}
\newcommand{\dytio}{Dy$_2$Ti$_2$O$_7$}
\newcommand{\wand}{WAND$^2$}

\def\equationautorefname~#1\null{Eq. (#1)\null}

\newcommand{\appref}[1]{\hyperref[#1]{Appendix~\ref*{#1}}}
\usepackage{physics}
\renewcommand\vec{\mathbf}

\usepackage{stackengine}
\stackMath

\begin{document}
\preprint{APS/123-QED}
%
%
\title
{Dipolar Spin Ice Regime Proximate to an All-In-All-Out N\'{e}el Ground State \\
in the Dipolar-Octupolar Pyrochlore Ce$_2$Sn$_2$O$_7$}
\author{D. R. Yahne}
\email[Corresponding author: ]{danielle.yahne@psi.ch}
\altaffiliation{Present address: Laboratory for Neutron and Muon Instrumentation, Paul Scherrer Institute, Villigen PSI, Switzerland}
\affiliation{Department of Physics, Colorado State University, 200 West Lake Street, Fort Collins, Colorado 80523-1875, USA}
\author{B. Placke}
\affiliation{Max Planck Institute for Physics of Complex Systems, N\"{o}thnitzer Strasse 38, Dresden 01187, Germany}
\author{R. Sch\"{a}fer}
\affiliation{Max Planck Institute for Physics of Complex Systems, N\"{o}thnitzer Strasse 38, Dresden 01187, Germany}
\author{O. Benton}
\affiliation{Max Planck Institute for Physics of Complex Systems, N\"{o}thnitzer Strasse 38, Dresden 01187, Germany}
\author{R. Moessner}
\affiliation{Max Planck Institute for Physics of Complex Systems, N\"{o}thnitzer Strasse 38, Dresden 01187, Germany}
\author{M. Powell}
\affiliation{Department of Chemistry, Clemson University, Clemson, South Carolina 29634-0973, USA}
\author{J. W. Kolis}
\affiliation{Department of Chemistry, Clemson University, Clemson, South Carolina 29634-0973, USA}
\author{C. M. Pasco}
\affiliation{Materials Science and Technology Division, Oak Ridge National Laboratory, Oak Ridge, Tennessee 37831, USA}
\author{A. F. May}
\affiliation{Materials Science and Technology Division, Oak Ridge National Laboratory, Oak Ridge, Tennessee 37831, USA}
\author{M. D. Frontzek}
\affiliation{Neutron Scattering Division, Oak Ridge National Laboratory, Oak Ridge, Tennessee 37831, USA}
\author{E. M. Smith}
\affiliation{Department of Physics and Astronomy, McMaster University, Hamilton, Ontario, L8S 4M1, Canada}
\affiliation{Brockhouse Institute for Materials Research, McMaster University, Hamilton, Ontario, L8S 4M1, Canada}
\author{B. D. Gaulin}
\affiliation{Department of Physics and Astronomy, McMaster University, Hamilton, Ontario, L8S 4M1, Canada}
\affiliation{Brockhouse Institute for Materials Research, McMaster University, Hamilton, Ontario, L8S 4M1, Canada}
\affiliation{CIFAR, MaRS Centre, West Tower 661 University Avenue, Suite 505, Toronto, Ontario, M5G 1M1, Canada}
\author{S. Calder}
\affiliation{Neutron Scattering Division, Oak Ridge National Laboratory, Oak Ridge, Tennessee 37831, USA}
\author{K. A. Ross}
\affiliation{Department of Physics, Colorado State University, 200 West Lake Street, Fort Collins, Colorado 80523-1875, USA}
\affiliation{CIFAR, MaRS Centre, West Tower 661 University Ave., Suite 505, Toronto, Ontario, M5G 1M1, Canada}
\date{\today}

%
\begin{abstract}
The dipolar-octupolar (DO) pyrochlores, R$_2$M$_2$O$_7$ (R = Ce, Sm, Nd), are key players in the search for realizable novel quantum spin liquid (QSL) states as a large parameter space within the DO pyrochlore phase diagram is theorized to host QSL states of both dipolar and octupolar nature. New single crystals and powders of \cesno, synthesized by hydrothermal techniques, present an opportunity for a new characterization of the exchange parameters in \cesno\ using the near-neighbor $XYZ$ model Hamiltonian associated with DO pyrochlores. Utilizing quantum numerical linked cluster expansion fits to heat capacity and magnetic susceptibility measurements, and classical Monte Carlo calculations to the diffuse neutron diffraction of the new hydrothermally grown \cesno\ samples, we place \cesno's ground state within the ordered dipolar all-in-all-out (AIAO) N\'{e}el phase, with quantum Monte Carlo calculations showing a transition to long-range order at temperatures below those accessed experimentally. Indeed, our new neutron diffraction measurements on the hydrothermally grown \cesno\ powders show a broad signal at low scattering wave vectors, reminiscent of a \textit{dipolar} spin ice, in striking contrast from previous powder neutron diffraction on samples grown from solid-state synthesis, which found diffuse scattering at high scattering wave vectors associated with magnetic {\it octupoles} and suggested an octupolar quantum spin ice state. We conclude that new hydrothermally grown \cesno\ samples host a finite-temperature proximate dipolar spin ice phase, above the expected transition to AIAO N\'{e}el order.

\end{abstract}
\pacs{xx.xx.mm}
\maketitle

\indent

\section{Introduction}
In $f$-electron systems, even apparently simple pseudospin-1/2 magnets can exhibit intriguing single-ion symmetry properties: beyond the ``usual'' case of three \textit{dipolar} spin components, some components may either be time-reversal symmetric quadrupoles, or transform as octupoles. These multipolar moments, especially when combined with magnetic frustration, can lead to a wealth of exotic phenomena. Of particular interest are ``hidden'' orders \cite{Kiss_URu2Si2,Santini_MultipolarReview,Kuramoto_MultipolarReview,Chandra_URu2Si2} elusive to common experimental probes, and long-range entangled quantum spin liquids (QSL).

The theoretical description of these systems naturally goes along with sets of new model Hamiltonians, which have indeed been found to host interesting phases, although the balance between the competing interactions which stabilize these phases can be delicate. 
In fact, frustrated materials commonly find themselves near phase boundaries between different orders, or proximate to exotic phases \cite{MPCYTO,Hallas2016,Hallas2017,MPCPyrochlore,Petit_ErSnO,Sarkis_YGO,MultiphaseMagnetismYTO}.  The possibility of tuning across these phase boundaries through chemical pressure, applied external pressure, or disorder makes the $f$-electron materials attractive systems for revealing unconventional physics. 

To advance our understanding of these multipolar systems, it is therefore imperative to develop a good microscopic description of relevant real materials, with respect to both the chemical properties of the samples under investigation and the appropriate Hamiltonian for their theoretical analysis. 

One ideal set of materials to develop such understanding is the dipolar-octupolar (DO) pyrochlores, R$_2$M$_2$O$_7$, where R$^{3+} = $\{Ce, Sm, Nd\} and M$^{4+}$ can be a range of transition metal ions, as they host a wealth of exotic phenomena due to their canonical frustrated geometry, range of interactions, single-ion anisotropies, and the ability to map to a simple pseudospin-1/2 system. In these 4$f$ rare-earth magnets,  the hierarchy of energy scales is headed by Coulomb interactions and spin-orbit coupling, followed by crystal electric field (CEF) effects from the surrounding charged environment, followed by intersite exchange interactions. It is the CEF that breaks the $(2J+1)$ degenerate free-ion ground state and leads to a ground state doublet that is classified by how it transforms under the point group symmetry of the rare-earth site ($D_{3d}$) and time-reversal symmetry~\cite{RauGingrasAnnuRevCMP}. When the  ground state CEF doublet is described by a pseudospin-1/2 in which different components transform as magnetic dipoles and octupoles, respectively, it is referred to as a dipole-octupole (DO) doublet \cite{Huang2014}. 

\begin{figure}[t!]
\includegraphics[scale = 0.9]{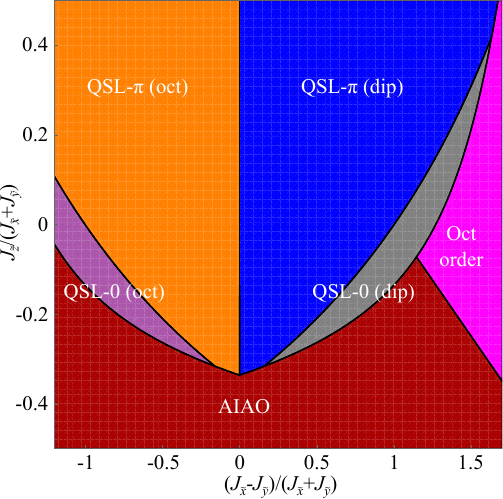}
\caption{ Ground state phase diagram of the dipolar-octupolar exchange Hamiltonian. Phase boundaries are based on the calculations from Benton~\cite{Benton2020}, which find four U$(1)$ QSL phases, labeled QSL-0 or $\pi$ (dip or oct), alongside dipolar and octupolar AIAO N\'eel phases. The dip or oct distinction refers to whether the dominant pseudospin component transforms as a dipole or an octupole, while the 0 or $\pi$ refers to the flux through the hexagonal plaquettes within the pyrochlore structure. The strong potential for realizable QSL materials is a huge motivation to further explore the DO pyrochlores in detail. \label{fig_pd_nomat}}
\end{figure}

The symmetry of the DO pyrochlores constrains the nearest-neighbor Hamiltonian to take the form of a simple $XYZ$ Hamiltonian,

\begin{equation}
\begin{aligned}
    H_{\mathrm{XYZ}} = 
    &\sum_{\langle i,j \rangle}
    \left[J_{\Tilde{x}}S_i^{\Tilde{x}}S_j^{\Tilde{x}}
    + J_{\Tilde{y}}S_i^{\Tilde{y}}S_j^{\Tilde{y}}
    + J_{\Tilde{z}}S_i^{\Tilde{z}}S_j^{\Tilde{z}}\right] \\
    &- g_z\mu_B \sum_i \vec{h} \cdot \hat{\vec{z}}_i
    \left(S_i^{\Tilde{z}}\cos\theta + S_i^{\Tilde{x}}\sin\theta\right),
\end{aligned}
\label{eq:Ham}
\end{equation}

\noindent where $S^{\Tilde{\alpha}}$ are the pseudospin components in the local $\Tilde{\alpha} = \Tilde{x}, \Tilde{y}, \Tilde{z}$ coordinate frame (related to the local $x,y,z$ coordinate frame, by a rotation of $\theta$ about the $y$ axis), $g_z$ is the $g$ factor, $\vec{h}$ is the magnetic field, and $\hat{\vec{z}}_i$ is the local anisotropy axis. The ground state phase diagram for the DO pyrochlores has been investigated in detail ~\cite{Li2017, Yao2020, Benton2020, Patri2020}, and the best current understanding is presented in \autoref{fig_pd_nomat}. It is clear and encouraging that this class of materials has huge potential to realize QSL states of both dipolar and octupolar character, as they occupy a large region of parameter space, alongside dipolar and octupolar all-in-all-out (AIAO) N\'{e}el phases. 

Within this exciting class of materials, the 4$f^1$ Ce$^{3+}$ family in particular has attracted intrigue as experimentally realizable QSL candidates~\cite{Sibille2015, Sibille2019, Gaudet2019, Gao2019, Smith2021}. Experimental work on \cezro, including extensive neutron scattering, heat capacity, and magnetic susceptibility, combined with theoretical calculations have constrained \cezro\ to lie within the U$(1)_\pi$ QSL state at low temperatures, on a boundary between octupolar and dipolar character, although the exact nature of the ground state is under debate~\cite{Smith2021,Bhardwaj2022,Desrochers2022}. Similarly, much recent interest was generated toward \cesno\ when neutron diffraction measurements on powder samples grown by solid-state synthesis found broad diffuse scattering at high scattering wave vectors ($|Q| > 5$ \AA$^{-1}$), attributed to scattering from an octupolar spin ice state or from higher order multipoles~\cite{Sibille2019,LOVESEY}.

Here we present an unexpected and remarkable result obtained by comparison of detailed quantum numerical linked cluster expansion (NLCE) calculations and Monte Carlo simulations to new \cesno\ single crystal and polycrystalline samples grown by hydrothermal synthesis techniques. Our work provides a constrained parametrization of a model Hamiltonian that allows the results for \cesno\ presented here to be placed on a phase diagram and contrasted with previous measurements of \cesno, as well as the related \cezro. Contrary to the previously reported \textit{octupolar} spin ice state, we report that a \textit{dipolar} all-in-all-out ground state is appropriate for \cesno, with a spin ice regime above the AIAO ordering. New diffuse neutron scattering measurements show a broad signal at low scattering wave vectors {\it only} (peaked near $|Q| \sim 0.6$ \AA$^{-1}$), and no evidence of higher-$Q$ scattering indicative of octupolar physics, consistent with the newly constrained ground state.

\begin{figure*}[ht!]
\includegraphics[scale = 1.0]{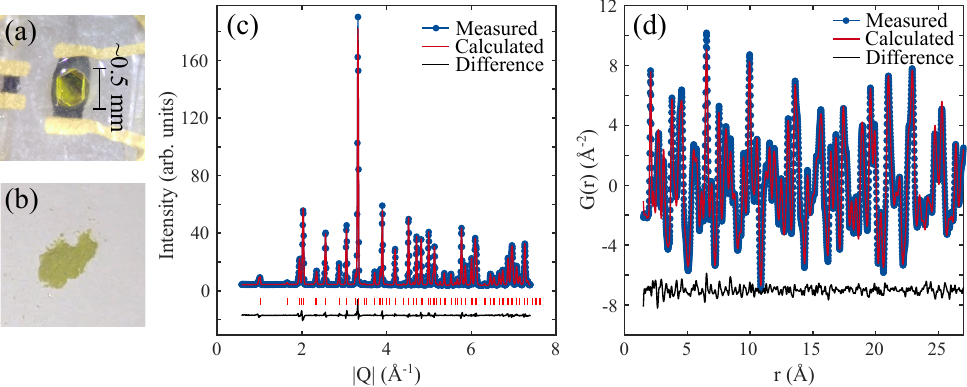}
\caption{ Representative (a) single crystal and (b) powder \cesno\ samples, both grown using hydrothermal techniques, used for heat capacity and magnetization measurements. Samples remained a bright yellow-green in air, with no visible signs of oxidation effects over time. (c) Rietveld refinement of room temperature powder neutron diffraction on beam line HB-2A. Data (blue points), calculated pattern (red line), difference pattern (black line) and Bragg positions of \cesno\ (red tick marks) are shown. Refined parameters can be found in \autoref{cesno_tab:hb2a}. (d) Pair distribution function (PDF) analysis of \cesno\ from total neutron scattering data taken on the NOMAD beam line at room temperature. The fitted curve (red) shows good agreement with the data (blue), finding no evidence of local distortions in the sample.  \label{cesno_fig:reitveld}}
\end{figure*}

\section{Sample synthesis and quality}

Ground state selection in several pyrochlore magnets is known to be sensitive to disorder.  Perhaps the best-known example is associated with \ybtio\ \cite{YTO_SC_vs_Powder, Ross2012, YTO_unconventional, YTO_stoich, YTO_AFM, YTO_pointdefects}. Ce-based pyrochlores are likely candidates for such sensitivity because, as with other light rare-earth ions, Ce is stable in both the 3+ and 4+ oxidation states. Gaudet \textit{et al.} showed that powder samples of \cezro\ grown by solid-state synthesis oxidize in air in a matter of hours\cite{Gaudet2019}. Unlike the rest of the magnetic rare-earth series, however, this type of defect is relatively benign, as Ce$^{4+}$ is $4f^0$ and consequently nonmagnetic. Nevertheless, a detailed characterization and understanding of Ce-based pyrochlores grown by different techniques is clearly a crucial open issue for this branch of topological materials physics. Therefore, we briefly outline the hydrothermal synthesis method and characterization of the sample quality. 

\subsection{Hydrothermal synthesis of \cesno}

Single crystal and powder samples of \cesno\ were grown via a hydrothermal synthesis method~\cite{Powell2019} [\autoref{cesno_fig:reitveld} (a) and (b)]. The advantage of hydrothermal synthesis, from a sample quality standpoint, is the significantly lower temperature required ($700^{\circ}$C), as compared with solid-state synthesis (typically $>1000^{\circ}$C), avoiding the temperatures at which tin oxide becomes volatile, thereby ideally minimizing oxidation. Approximately $13$~g of the pyrochlore stannate powder was prepared for neutron measurements using a mixture of $30-50\%$ SnO and $70-50\%$ SnO$_2$ in order to reduce any residual \cefo\ to \ceth, thus minimizing external impurities. 

We expect negligible stuffing of the A-site \ceth\ on the B-site \snfo\ due to the large size difference ($1.01$ versus $0.69$~\AA), unlike Yb and Ti ($0.87$ versus $0.61$~\AA) in \ybtio \cite{Ross2012}. If present, such stuffing would create a corresponding defect in the oxygen sublattice, leading to a noticeable change in color. As shown in \autoref{cesno_fig:reitveld} (a) and (b), the single crystal and polycrystalline samples are bright yellow with a green tint, and we observe no evidence of oxidation effects over time while in air, in stark contrast to the sister compound \cezro\ which turns black on the order of hours. Furthermore, we have collected temperature-dependent, full-sphere single crystal x-ray datasets at $100$, $200$, and $300$~K. The oxygen atom positions were refined on lower symmetry sites and we see no flattening or movement of the thermal anisotropic displacement parameters, suggesting an absence of defects in the oxygen sublattice at any measurable concentration. The refined lattice parameter of $10.6464(4)$~\AA, within $0.01$~\AA\ of Tolla \textit{et al.}~\cite{Tolla1999} and Sibille \textit{et al.}~\cite{Sibille2019}, also suggests negligible oxidation. The remaining question is whether there is evidence of \cefo\ stuffing on \snfo\ sites, which is possible given the comparable ionic radii ($0.87$ versus $0.69$~\AA). Our x-ray refinement places an upper bound of $3\%$ for such B-site stuffing of \cefo.

\subsection{Room temperature powder neutron diffraction (PND) and analysis}

Room temperature powder neutron diffraction (PND) measurements were performed on beam lines HB-2A at the High Flux Isotope Reactor (HFIR) and NOMAD at the Spallation Neutron Source (SNS) at Oak Ridge National Laboratory (ORNL)~\cite{Calder2018}. Room temperature PND on HB-2A [\autoref{cesno_fig:reitveld} (c)] utilized a Ge(115) monochromator with an incident wavelength of $1.54$~\AA.  The resulting powder neutron diffraction pattern was refined using the \textsc{Fullprof} software suite which implements the Rietveld refinement method~\cite{Fullprof_Rodriguez}. We find a lattice parameter of $10.64542(5)$~\AA, in agreement with our single crystal x-ray diffraction. A small nonmagnetic impurity of CeO$_2$ is present, and makes up $\sim 3\%$ of the powder material by weight. 

The total scattering measurement on NOMAD, with a $Q_{\mathrm{max}} = 40$ \AA$^{-1}$, allows an atomic pair distribution function (PDF) analysis to be carried out to further constrain the local structure of the hydrothermally grown sample [\autoref{cesno_fig:reitveld} (d)]. The PDF analysis gives an atom-by-atom histogram of all of the pair-pair correlations in a material as a function of real-space distance $r$. The data were processed through the NOMAD autoreduction software and we utilized \textsc{PDFgui} \cite{Farrow_2007} to perform the PDF refinement. This analysis resulted in a similar lattice parameter as was found using HB-2A, and there was no evidence of local structure distortions.

\begin{figure}[t!]
\includegraphics[scale = 1.0]{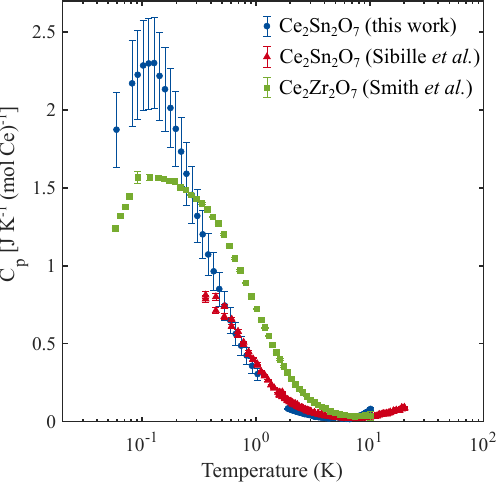}
\caption{Single crystal (T$<1$~K) and powder (T$>1$~K) heat capacity on hydrothermally synthesized \cesno\ compared to previously published polycrystalline data by Sibille \textit{et al.} \cite{Sibille2019} as well as \cezro\ from Smith \textit{et al.}\cite{Smith2021}. Note that the data presented in this work are the total heat capacity, while the comparative works present the isolated magnetic contribution. However, the phonon contribution below $10$~K is negligible and thus does not affect the conclusions of this work. We find good agreement with Sibille \textit{et al.}, and the shift of the high temperature tail to lower temperatures when compared to \cezro\ suggests a smaller scale of the intersite interaction strengths. \label{fig_specificheat}}
\end{figure}


\section{Estimating the exchange parameters in the spin Hamiltonian for \cesno}

\subsection{Heat capacity and susceptibility}

Heat capacity measurements were performed on a $0.26 \pm 0.03$~mg single crystal of \cesno\ (\autoref{fig_specificheat}) from $T = 1$~K down to $T = 0.05$~K using a Quantum Design PPMS with dilution insert. Similar low temperature measurements were carried out on hydrothermally grown powder samples, with additional measurements taken above $1.9$~K. The large uncertainty ($\sim 10\%$) at low temperatures is attributed to the extremely small sample mass, a contribution that is typically ignored, and is determined by multiple successive weight measurements. Note that this uncertainty is associated with the overall \textit{scale} of the heat capacity; the relative uncertainty, data point to data point at different temperatures, is much smaller.

Our results above $0.5$ K are in good agreement with the literature~\cite{Sibille2015,Sibille2019}.  We find no evidence for a transition to long-range order down to the lowest measured temperatures, and instead observe a broad peak in the heat capacity, the shape of which is similar to that of the sister compound \cezro~\cite{Smith2021,Gao2019}. Importantly, we note the high temperature tail of the heat capacity peak is shifted to lower temperature compared to \cezro, indicating that  \cesno\ hosts comparatively smaller intersite interaction strengths, as we will show in \autoref{main_NLCE}.

Susceptibility measurements were also carried out using a Quantum Design MPMS3 with iHelium3 insert. Powder and single crystal samples were mounted to straws using grease and measured with an applied field of $H = 0.01$~T (field aligned along $[111]$ for single crystals). Results are shown in \autoref{fig_Susc} (a).

\begin{figure}[b!]
\includegraphics[scale = 0.95]{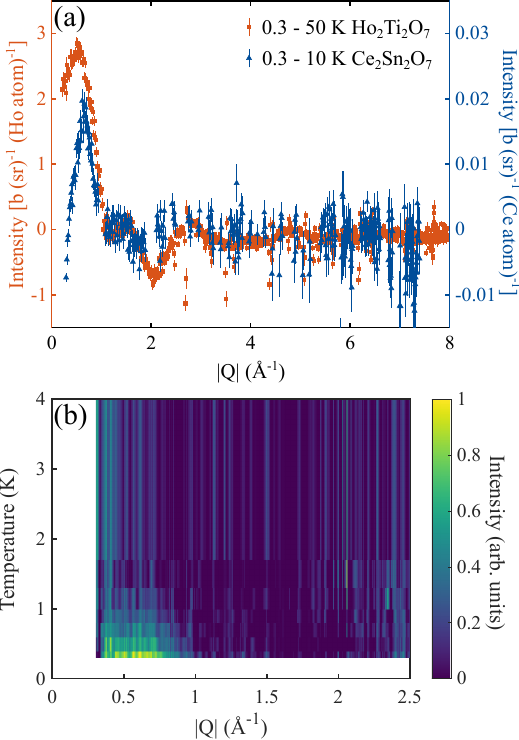}
\caption{ (a) High temperature subtracted powder diffuse neutron scattering of the known dipolar spin ice compound \hotio\ compared with newly synthesized \cesno\ in absolute units. Both materials show a broad diffuse peak at $|Q| \sim 0.6$ \AA$^{-1}$. (b) Temperature dependence of the diffuse scattering of \cesno\ with paramagnetic background removed. The intensity increase near $|Q| \sim 0.6$~\AA$^{-1}$ is seen to onset around $1.5$ K. \label{fig_hotio}}
\end{figure}


\subsection{Low temperature neutron diffraction}

We collected low temperature PND on the \wand\ beam line at HFIR (ORNL) using a Ge(113) monochromator for an incident wavelength of $1.488$~\AA~(E$_i \sim 37$ meV), and a closed-cycle refrigerator. These measurements on the hydrothermally grown \cesno\ samples were performed as a function of temperature from a base temperature of $T = 0.3$~K, well below the temperature at which octupolar correlations are expected $\sim 1$~K, up to $10$ K.  The high flux and $^3$He position-sensitive detector makes \wand\ an ideal instrument for probing weak, diffuse neutron diffraction signals, as may be expected from octupolar scattering at high $Q$. The diffuse scattering signal is isolated by performing a high temperature subtraction of the paramagnetic background $10$ K dataset.  This protocol allowed us to put the net diffuse neutron scattering on an absolute scale, to compare with earlier diffuse scattering measurements performed on \cesno\ powder samples grown by solid-state synthesis. However, neither work is normalized using a vanadium standard and are therefore susceptible to uncertainties associated with absorption effects. This has no bearing on the difference in the $Q$ dependence of the diffuse neutron scattering signals.

Intriguingly, our new neutron diffraction results show the diffuse scattering to peak at small $|Q| \sim$ 0.6 \AA$^{-1}$, and no diffuse scattering above background was observed beyond $\sim 2$ \AA$^{-1}$ [\autoref{fig_hotio} (a)].  The temperature dependence of the diffuse scattering from the new hydrothermally grown \cesno\ powder sample is shown in \autoref{fig_hotio} (b). These data show that the low-$Q$ diffuse scattering onsets near $1.5$~K where we also see the initial increase in the heat capacity.  This diffuse scattering was further corroborated by similar low temperature measurements on beam line HB-2A at $T=0.3$~K and $T=4$~K.

This new diffuse neutron scattering from \cesno\ strongly resembles that from the canonical dipolar spin ice pyrochlore \hotio. We compare the diffuse scattering results from these two pyrochlore powders in \autoref{fig_hotio} (a) (see Appendix B for experimental details). In both cases, this diffuse scattering is inelastic scattering that has been integrated in the diffraction measurement, as the incident energy is much larger than the bandwidth of inelastic scattering known to be relevant to \cesno\ and \hotio\cite{Sibille2019,Clancy2009}. The dipole moments associated with Ho$^{3+}$ in \hotio\ are about 8 times larger than those associated with \ceth\ in \cesno; hence the overall intensity of the diffuse dipolar spin ice diffraction pattern is much stronger in \hotio\ compared with the new \cesno\ sample.  Nonetheless, the qualitative similarity between the two sets of diffuse scattering is clear.

\begin{figure}[t!]
\includegraphics[scale = 1.0]{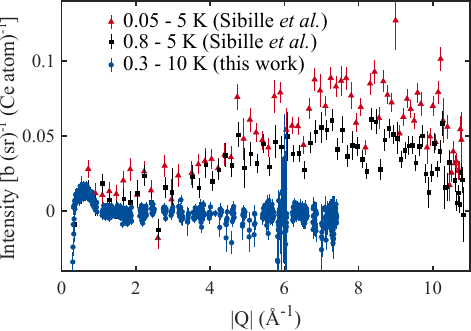}
\caption{ Diffuse neutron scattering from hydrothermally grown \cesno\ at $0.3$~K taken on \wand\ is shown as the blue circles. All data have been converted to absolute units, and paramagnetic background has been removed. We compare the \cesno\ diffuse scattering to high-$Q$ diffuse scattering found by Sibille \textit{et al.}~\cite{Sibille2019} from samples grown by solid-state synthesis. We do not see evidence of octupolar scattering at high $Q$, and instead find subtle diffuse scattering akin to a dipolar spin ice [see \autoref{fig_hotio} (a)].  \label{fig_diffuse}}
\end{figure}

The diffuse scattering of both \cesno\ samples grown by hydrothermal and solid-state synthesis techniques is directly compared in \autoref{fig_diffuse}. The two works strongly contrast, with the earlier \cesno\ measurements showing a much increased diffuse scattering beyond $|Q| \sim 3$ \AA$^{-1}$, peaking at $|Q| \sim 8$ \AA$^{-1}$ with nearly a factor of 4 times the intensity than observed at low $Q$. That is, it is hard to imagine that the diffuse scattering at low temperature could be more different between the new hydrothermally grown \cesno\ powder and the previously studied \cesno\ powder sample grown by solid-state synthesis.

\subsection{Numerical linked cluster expansions and Monte Carlo calculations and their comparison to experiment}
\label{main_NLCE}

\begin{figure}[t!]
\includegraphics[scale = 1.0]{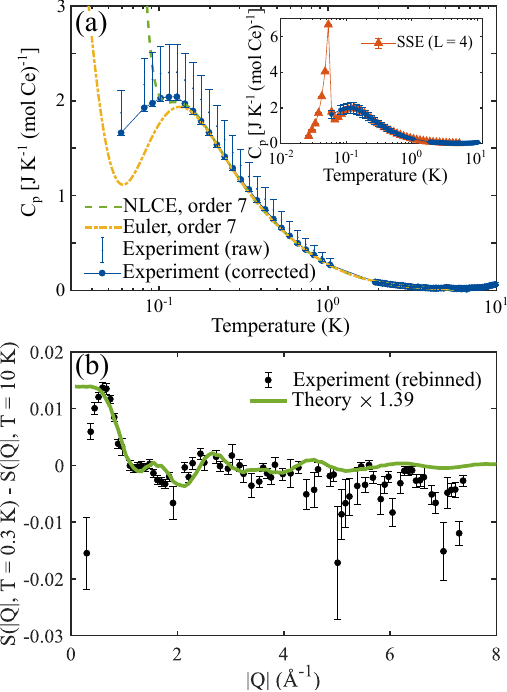}
\caption{ (a) The measured \cesno\ heat capacity, along with appropriate best fits to the NLCE theory up to order 7. The experimental heat capacity is corrected by the fitted parameter $\varepsilon_c$ to account for the systematic mass uncertainty. Inset: the full QMC calculated heat capacity beyond experimentally accessible temperatures using parameters near the best fit parameters (see Appendix G). This finds a sharp anomaly signifying an expected phase transition to the AIAO state near $0.04$~K.  (b) The measured $S(|q|)$ for hydrothermally grown \cesno\ along with the theoretical calculation using best fit parameters. This shows both C$_\mathrm{p}$ and $S(|q|)$ can be well described by  calculations based on \autoref{eq:Ham}, and allow us to determine our estimate for the terms in the exchange Hamiltonian (except for $\theta$). \label{fig_NLC}}
\end{figure}


\begin{figure}[t!]
\includegraphics[scale = 1.0]{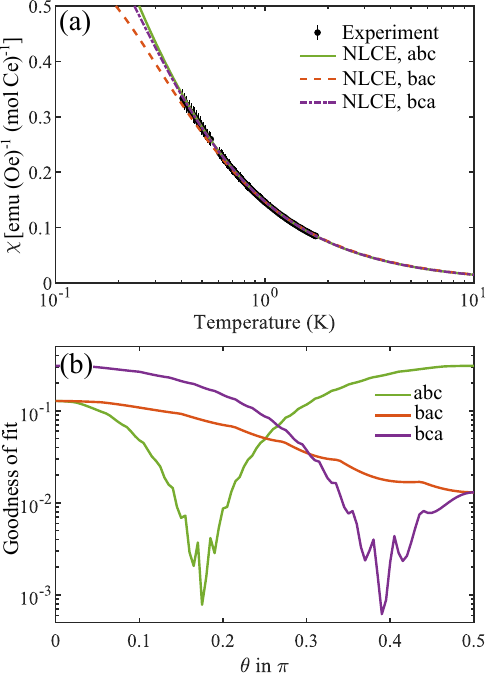}
\caption{ (a) Measured magnetic susceptibility $\chi(T)$ with an applied field strength of $0.01$~T, and NLCE fits that establish $\theta \sim 0.2\pi$ in the $\{a,b,c\}$ permutation. (b) The goodness of fit to the susceptibility data is shown as a function of $\theta$, and we find a strong minimum (i.e. strong constraint) near $\theta \sim 0.2\pi$. While a better goodness of fit is found for the $\{b,c,a\}$ permutation (which would constrain $\theta \sim 0.4\pi$), this permutation does not result in a reasonable $S(|q|)$ fit.  \label{fig_Susc}}
\end{figure}


To understand the underlying behavior of \cesno, we invoke theoretical calculations that allow a parametrization of the spin Hamiltonian given in \autoref{eq:Ham}. We utilize a quantum numerical linked cluster expansion of the specific heat and susceptibility as well as classical, and wherever possible quantum, Monte Carlo simulations of the powder-averaged neutron structure factor to determine the coefficients of the nearest-neighbor exchange interactions and the corresponding ground state, as well as finite-temperature properties of \cesno. At finite temperatures, NLCE provides an opportunity to calculate observables in the thermodynamic limit without parameter restrictions plagued by other numerical methods (see App. D), enabling parametrization throughout the whole DO pyrochlore phase space.

The procedure is similar to the one used to constrain \cezro\ to lie within the U$(1)_{\pi}$ QSL ground state \cite{Smith2021} and is described in detail in Appendix D as well as briefly outlined here. First, we rewrite \autoref{eq:Ham} in terms of $J_a$, $J_b$, and $J_c$, where $\{a,b,c\}$ are simply a permutation of $\{\Tilde{x},\Tilde{y},\Tilde{z}\}$ such that $|J_a|>|J_b|,|J_c|$ and $J_b \geq J_c$. We compute the specific heat for the full range of possible values of $J_b/J_a$ and $J_c/J_a$, optimizing the energy scale, $J_a$, in each case using the  high temperature tail of the experimental data $C(T)$. An additional fitting parameter, $\varepsilon_c \in [-1,1]$, is included in the fits to the heat capacity data to account for the systematic mass uncertainty from the extremely small samples. Using the estimated energy scales, we then compute the susceptibility $\chi(T)$ and structure factor, $S(|q|, T = 0.3~K) - S(|q|, T = 10~K)$, for each value of $J_{a,b,c}$, each value of $\theta$, and each permutation of exchange parameters $\left(\{\Tilde{x},\Tilde{y},\Tilde{z}\} = \{a,b,c\}, \{b,a,c\}, \{b,c,a\}\right)$ (see Appendix D), optimizing the value of the $g$ factor $g_z$ for each case. Finally, we minimize the total goodness of fit $\chi^2$ to find the set of exchange parameters and permutation that best agrees with the experimental data.

Our best fit is achieved with the following exchange parameters in meV: 
\begin{equation}
\begin{aligned}
    (J_{\Tilde{x}},J_{\Tilde{y}},J_{\Tilde{z}}) &= (0.045, -0.001, -0.012), \\
    \theta &= 0.19\pi, \ \  g_z = 2.2.
\end{aligned}
\label{eq:Js}
\end{equation}

\noindent This parametrization gives excellent agreement with the experimental heat capacity data down to $0.1$~K [\autoref{fig_NLC} (a)], including the fit to the systematic mass uncertainty $\varepsilon_c$. The fit to the magnetic susceptibility strongly constrains the value of mixing angle $\theta$. While two permutations $\left(\{a,b,c\}\ \mathrm{and}\ \{b,c,a\}\right)$ result in minima in the goodness of fit [\autoref{fig_Susc} (b)], only the $\{a,b,c\}$ permutation in \autoref{eq:Js} produced a reasonable $S(|q|)$ fit [\autoref{fig_NLC} (b)].

\begin{figure}[t!]
\includegraphics{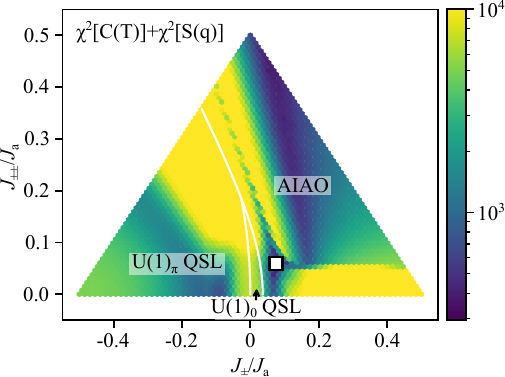}
\caption{Map of the goodness of fit ($\chi^2$) in the reduced phase diagram that shows the best fit parameter location (white square) of hydrothermally grown \cesno. \label{fig_bestfit}}
\end{figure}

Importantly, these parameters place the ground state of \cesno\ within the \textit{ordered dipolar AIAO} region of the phase diagram, proximal to the U$(1)_0$ QSL phase, as seen in \autoref{fig_bestfit}. This suggests a phase transition to AIAO order will occur at low temperatures, although we emphasize that we observe no direct evidence for this in experiment.

As our best fit estimates for the terms in the spin Hamiltonian place \cesno\ in a nonfrustrated region of the DO phase diagram, quantum Monte Carlo (QMC) calculations can be carried out without a sign problem using these parameters. Our QMC simulation using the parameters in \autoref{eq:Js} predicts a phase transition temperature just within the temperature range of our experiments (see Appendix G). However, the transition temperature is very sensitive to small variations of the Hamiltonian parameters in this region of the phase diagram. For example, a small modification of the parameters to $(J_{\tilde{x}}, J_{\tilde{y}}, J_{\tilde{z}})=(0.0457, -0.0014, -0.0110)~{\rm meV}$ produces a transition temperature below the experimental range and therefore consistent with experiment, without significantly reducing the agreement with other quantities. A comparison of the experimental data with the QMC results for the slightly modified parameter set is shown in the inset of \autoref{fig_NLC} (a). 

It is worth noting that for the dipolar, classical spin ices, \hotio\ and \dytio, classical Monte Carlo simulations with realistic interaction parameters show a phase transition at very low temperatures from a classical, disordered dipolar spin ice state to an ordered spin ice state\cite{Melko2001}.  However, the temperature at which this predicted phase transition occurs is also very sensitive to the precise Hamiltonian parameters employed.  This predicted phase transition has yet to be observed experimentally in either \hotio\ or \dytio, despite numerous attempts.

\begin{figure}[hb!]
\includegraphics[width=\columnwidth]{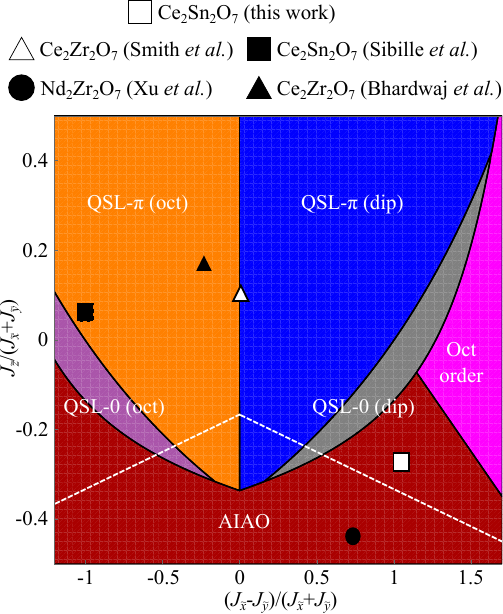}
\caption{Ground state phase diagram of the dipolar-octupolar exchange Hamiltonian, with parameter estimates for various materials \cite{Sibille2019, Xu2019, Bhardwaj2022, Smith2021}. Here the phase diagram is shown without the reduction of parameter space used in \autoref{fig_bestfit} so that dipolar and octupolar phases are distinguished. Phase boundaries are based on the calculations from Benton~\cite{Benton2020}. The dashed lines represent the classical phase boundary of the AIAO order. The parameter estimate for \cesno\ from the present work falls within the AIAO phase, in a region where the ordering is classically unstable but stabilized by quantum fluctuations for $S=1/2$.}
\label{fig_pd_materials}
\end{figure}


\section{Discussion}

While the protocol for determining \autoref{eq:Js} is already outlined, it is worthwhile to describe how this combination of heat capacity, magnetic susceptibility and diffuse neutron scattering measurements \textit{strongly constrains} the best fit Hamiltonian parameters.  First, the high temperature tail of the heat capacity constrains the overall scale of exchange interactions, while the height and shape of the smooth peak of the heat capacity constrain the ratios.  In particular, a strongly negative value of $J_{\pm}$ leads to a rather lower heat capacity peak, as observed in Ce$_2$Zr$_2$O$_7$. On the other hand, a sufficiently large positive $J_{\pm}$ leads to a sharp phase transition without any preceding smooth peak at higher temperature. The presence of  smooth peak with height $\gtrsim 2.0\ {\rm J~K^{-1}~mol^{-1}}$, thus forces a relatively small value of $|J_{\pm}|/J_a$. The shape of the curve on the approach to the peak, is further constraining, and narrows the region of good fits to the region around our optimal parameter set.

The permutation of parameters $\{ J_a, J_b, J_c \} \to \{ J_{\tilde{x}}, J_{\tilde{y}}, J_{\tilde{z}} \}$ is not fixed by the heat capacity but can be fixed using neutron scattering and susceptibility data.  In particular, the well-developed, spin-ice-like scattering at low $Q$ could only be reproduced in our simulations by allowing the largest exchange parameter to be one of the dipolar couplings $J_{\tilde{x}}$  or $J_{\tilde{z}}$. When all three datasets (heat capacity, neutron scattering, and susceptibility) are taken together, the fits are overconstrained rather than underconstrained, with a relatively small region of parameter space enabling acceptable agreement with all three.

With the NLCE calculations having established a dipolar AIAO ground state for \cesno, we interpret the measured diffuse neutron diffraction at $T = 0.3$~K as a spin ice regime above the AIAO transition. Further, it is interesting to note that the $\theta > 0$ term in the best fit Hamiltonian implies spin waves should be visible in nonzero magnetic field. Inelastic neutron scattering measurements could therefore corroborate the fitted Hamiltonian parameters, but are not presently straightforward due to the small sizes of available single crystals. Nonetheless, this would be an important future step after sample size optimization is developed.

\autoref{fig_pd_materials} shows the ground state phase diagram of the pyrochlore $XYZ$ Hamiltonian without the reduction of parameter space by permutations, in contrast to \autoref{fig_bestfit}, so that dipolar and octupolar phases are distinguished. Available parametrizations of the Hamiltonian for \cesno \cite{Sibille2019}, \cezro\ \cite{Smith2021, Bhardwaj2022} and the ordered dipolar-octupolar pyrochlore \ndzro \cite{Xu2019} are shown. Our estimate for the parameters of \cesno\ is far from the previous estimate from Sibille \textit{et al.}\cite{Sibille2019}, and also far from estimates for the sister material \cezro \cite{Smith2021, Bhardwaj2022}. This surprising degree of sensitivity of the exchange parameters to both the transition metal M and, at least potentially, to disorder, implies a high degree of tunability in Ce pyrochlores which could be used to further explore the full phase diagram. 

The phase diagram in \autoref{fig_pd_materials} includes some regions where quantum fluctuations substantially renormalize the phase boundaries from what one would obtain in a classical treatment of \autoref{eq:Ham}. This is illustrated by the white dashed lines in \autoref{fig_pd_materials}, which show the classical phase boundary of the AIAO antiferromagnetic phase. According to our parametrization, \cesno\ falls in a regime where, classically, one would expect a disordered spin ice ground state, but where quantum fluctuations stabilize antiferromagnetic AIAO order.  This then rationalizes our observation of a dipolar spin ice regime at temperatures above an expected phase transition to AIAO order.

\cesno\ thus appears as an example of a material where strong quantum effects favor ordering over disorder. Such order by disorder is a known phenomenon in frustrated magnets, but \cesno\ presents a striking example both because it is far from the regime where quantum effects can be treated perturbatively and because the resulting order is not selected from among the classical manifold of ground states.

\section{Summary and conclusions}

In summary, we report new thermodynamic, magnetic, and diffuse neutron scattering data from powder and small single crystals of \cesno\ grown by new hydrothermal techniques. Modeling of this experimental data using NLCE, classical Monte Carlo techniques and (where possible) quantum Monte Carlo techniques allow us to place \cesno\ in the region of the general $XYZ$ DO phase diagram where a dipolar AIAO ground state is selected.

Diffuse neutron scattering measurements performed on the new samples of \cesno\ grown by hydrothermal techniques possess a similar $Q$ dependence to that established  in the canonical dipolar spin ice \hotio.  These new measurements are distinctly different from the diffuse scattering observed at high $Q$ and previously ascribed to octupolar correlations in \cesno\ samples grown from solid-state synthesis. The origin of the difference between these new diffuse neutron results and those reported earlier is not understood.  On the one hand, we may expect different levels, and perhaps types, of weak disorder present in the same material synthesized using different techniques. Nevertheless, while the samples themselves could be different, the stark difference in low temperature diffuse neutron scattering would imply an extreme sensitivity to small levels of weak disorder. 

Our results, however, are internally consistent.  Modeling of our comprehensive experimental datasets places our new sample of \cesno\ in a region of the phase diagram where a dipolar AIAO ground state is selected, and the octupolar QSL states are not nearby.  Our new diffuse neutron scattering measurements are consistent with this, and can themselves be described by this same theory.  That is, for this dipolar AIAO region of the DO phase diagram, strong diffuse scattering at large $Q$ is not expected.  All of this supports our conclusion that for these new hydrothermally grown \cesno\ samples, we observe a proximate dipolar spin ice regime at finite temperature above a predicted dipolar AIAO ground state.

\section{acknowledgements}
The authors thank V. Por\'{e}e and R. Sibille for useful discussions and providing experimental data for comparison. We also thank J. Liu for performing the mail-in experiment at the NOMAD beam line at ORNL. D. R. Y. acknowledges useful discussions with J. Paddison. This research used resources at the High Flux Isotope Reactor and the Spallation Neutron Source at Oak Ridge National Laboratory, which was sponsored by the Scientific User Facilities Division, Office of Basic Sciences, U.S. Department of Energy. D. R. Y and S. C. acknowledge support from the U.S. Department of Energy, Office of Science, Office of Workforce Development for Teachers and Scientists, Office of Science Graduate Student Research (SCGSR) program. The SCGSR program is administered by the Oak Ridge Institute for Science and Education for the DOE under Contract No. DE-SC0014664. D. R. Y., K. A. R., M. P. and J. W. K. acknowledge funding from the Department of Energy Award No. DE-SC$0020071$. Specific heat and magnetization measurements by A. F. M. and C. M. P. were supported by the U.S. Department of Energy, Office of Science, Basic Energy Sciences, Materials Science and Engineering Division. This work was partly supported by the Deutsche Forschungsgemeinschaft under Grant No. SFB 1143 (Project No. 247310070) and the cluster of excellence ct.qmat (EXC 2147, Project No. 390858490). Additionally, this work was supported by the Natural Sciences and Engineering Research Council of Canada (NSERC).

\appendix

\section{Neutron Rietveld refinement}
\label{appendix:hb2a}

\begin{table*}[ht!]
\caption{Refined room temperature parameters for \cesno\ from neutron diffraction on HB-2A using the \textsc{Fullprof} Rietveld refinement.}
\scalebox{1.0}{
\begin{tabular}{cccccccccc}
\hline
\multicolumn{4}{l}{Lattice parameter (\AA): \ 10.64542(5)} & \multicolumn{2}{l}{Space group: \ Fd-3m} & \multicolumn{2}{l}{Temperature: \ 300 K} &   &  \\ \hline
Atoms       & $x$       & $y$       & $z$       & $B_{11}$      & $B_{22}$      & $B_{33}$      & $B_{12}$      & $B_{13}$      & $B_{23}$    \\
Ce ($16d$)    & 0.5    & 0.5    & 0.5     & 0.0005(1)   & 0.0005(1)   & 0.0005(1)   & -0.00031(10)   & -0.00031(10)   & -0.00031(10) \\
Sn ($16c$)    & 0       & 0      & 0      & 0.00019(10)   & 0.00019(10)   & 0.00019(10)   & -0.00002(9)   & -0.00002(9)   & -0.00002(9)  \\
O ($48f$)     & 0.3307(1)    & 0.125    & 0.125 & 0.00076(10)   & 0.00089(8)   & 0.00089(8)   & 0   & 0   & 0.00047(11)           \\
O ($8b$)      & 0.375     & 0.375     & 0.375     & 0.0009(2)     & 0.0009(2)     & 0.0009(2)     & 0             & 0             & 0           \\ \hline
\multicolumn{2}{l}{Lambda (\AA): \ 1.54} & & \multicolumn{2}{l}{$R_{\mathrm{Bragg}}$: \ 2.07}  &  \multicolumn{2}{l}{$R_f$: \ 1.34}  &   & &
\\ \hline
\end{tabular}}
\label{cesno_tab:hb2a}
\end{table*}

Room temperature neutron diffraction was performed on beam line HB-2A with an incident wavelength of $\lambda = 1.54$ \AA\ [\autoref{cesno_fig:reitveld} (b)] using a standard vanadium can. Rietveld refinement using the \textsc{Fullprof} software suite was utilized for fitting the lattice parameters, atomic displacement parameters (ADPs), and occupancy, and the refined values can be found in \autoref{cesno_tab:hb2a}. A nonmagnetic CeO$_2$ phase was necessary to account for some spurious peaks, and was found to make up less than $3\%$ of the sample by weight. Importantly, the addition of an oxygen atom at the $8a$ crystallographic site (the oxygen vacancy site in a perfect pyrochlore) did not result in an improved fit and the occupancy refinement oscillated around zero. Given the resolution of the measurement, we can obtain an upper bound of $\sim 2\%$ oxygen occupancy. Refined parameters are therefore found to be in agreement with PND from Sibille \textit{et al.}~\cite{Sibille2019} taken on the HRPT beam line at SINQ, Paul-Scherrer Institute.

Room temperature time-of-flight total scattering was performed on the NOMAD beam line, taking advantage of the mail-in program, in order to perform an atomic pair distribution function analysis. This analysis technique is sensitive to local (short-range) disorder effects and is complementary to traditional crystallographic analyses that probe the long-range average structure.

The returned data was carefully background subtracted and given in $S(|q|)$ and $G(r)$ for a variety of $Q_{\mathrm{max}}$ cutoff values, processed by the NOMAD auto-reduction software. A $Q_{\mathrm{max}} = 40$~\AA$^{-1}$ was chosen to obtain sufficient real-space resolution while minimizing the high-frequency noise from the high-$Q$ data. The \textsc{PDFgui} software was used to perform least-squared refinements of the modeled data with fitted lattice parameters, thermal ADPs, and the $48f$ oxygen position. The refined pattern agrees well with the experimental data and fitted lattice parameter of $10.6476$~\AA, and the lack of spurious peaks at expected vacancy positions suggests no local distortions are present [\autoref{cesno_fig:reitveld} (d)]. The residual found is comparable to that found from fitting a silicone standard under the same instrumental conditions.

\section{\hotio\ measurements and Data treatment}
\label{appendix:hotio}

Powder neutron diffraction was performed on canonical dipolar spin ice \hotio\ at the HB-2A beam line (HFIR, ORNL) using an incident wavelength of $1.54$\AA\ using the Ge$(115)$ monochromator. Powder samples were prepared using the conventional solid-state synthesis method\cite{Ghasemi2018,Brixner1964} and subsequently loaded into a standard aluminum can as a pressed pellet. Diffraction patterns were measured as a function of temperature from $0.3$ to $50$~K, both well below and above the temperature at which spin ice correlations emerge, around $2$~K.

In both low temperature PND measurements of \cesno\ and \hotio, high temperature background subtraction was utilized to isolate the pure magnetic signal at low temperatures. In the case of \cesno, Sibille \textit{et al.} find a temperature-independent, featureless paramagnetic scattering above $2$~K, justifying our choice of using $10$~K as paramagnetic background. \hotio\, however, has been shown to posses short-range correlations up to nearly $20$~K\cite{Bramwell2001}, and thus $50$~K was chosen for the paramagnetic background. The results of this subtraction are shown in Fig. 4; and we note that the nuclear Bragg peaks have been masked for clarity in both materials.

\section{Powder heat capacity measurements}
\label{appendix:hc}

For powder heat capacity measurements above $1.9$~K, a pellet was pressed from the sample powder, while measurements below $1.9$~K required a pellet made using an equal-mass combination of the sample powder and silver powder to enhance the thermal conductivity and sample coupling. The contribution of the silver to the specific heat was subtracted using in-house data for the specific heat capacity of silver. The relative contribution of silver decreases from $8\%$ at $1.9$~K, to $1\%$ at $1$~K, and drops below $0.1\%$ at $0.4$~K. After this careful background subtraction, the powder heat capacity data were found to be in good agreement with the single crystal measurements.

\section{Numerical linked cluster expansion}
\label{appendix:nlce}
Obtaining thermodynamic observables in three-dimensional frustrated materials, such as dipolar-octupolar pyrochlores, is extremely challenging as different numerical methods face different limitations.

First, the exponentially growing complexity limits exact diagonalization techniques such as finite-temperature Lanczos to small systems\cite{Schnack2018}. Second, finite-temperature methods based on tensor networks or matrix product states beyond one dimension fail to handle the entanglement and tend to develop biases toward ordered states\cite{Verstraete_matrix_2004,Feiguin_finite_2005}. Lastly, Monte Carlo approaches like stochastic series expansion (SSE), cf. Appendix G, are restricted to unfrustrated parameter regimes due to the infamous sign problem\cite{Kato2015,Huang2020}.

At finite temperature, the numerical linked cluster expansion \cite{Tang2013} opens up the possibility to calculate observables in the thermodynamic limit without any parameter restriction. Starting from a single tetrahedron, the algorithm generates clusters of finite size by building up the whole lattice in a systematic way. These clusters are then solved using exact diagonalization. The generic $XYZ$ Hamiltonian in \autoref{eq:Ham} does not preserve any continuous spin symmetry which makes numerical methods extremely challenging. Thus, we rely only on spatial symmetries which are determined and optimized in an automatic way\cite{Schaefer2020}.
The tetrahedral expansion of the pyrochlore lattice exhibits a large number of independent spatial symmetries.
The region of convergence ($T \gtrsim 0.1$~K) is comparable to temperatures used in the experiment and therefore allows a systematic analysis in the full parameter regime. A similar procedure was followed in our previous analysis of \cezro \cite{Smith2021}.

\subsection{Specific heat} 
Many thermodynamic observables like the specific heat or entropy depend only on the spectrum of the Hamiltonian. This can be exploited to reduce the possible parameter sets ($J_x$, $J_y$, $J_z$). First, the overall energy scale $\alpha$ can be neglected in numerical simulations as it simply adjusts the temperature unit of the specific heat curve $C(T)$. Adjusting the energy scale by $\alpha$ ($E_i\rightarrow E_i^\prime = \alpha E_i$) induces a shift of $\alpha^{-1}$ in the specific heat:
\begin{align}
    C(T)\rightarrow C^\prime(T) = C(T/\alpha)
\end{align}
Thus, the three-dimensional parameter space is reduced to two parameters. Second, ($J_x$, $J_y$, $J_z$) can be permuted by unitary rotations in spin space and the permutations do not affect the spectrum of the Hamiltonian for $ \vec{h} = \vec{0}$. Therefore, we can reduce the possible parameter sets to the characteristic triangular shape of possible exchange parameters shown in \autoref{fig_bestfit}.

Given a specific parameter set, we calculate the specific heat curve in sixth order in a tetrahedral expansion including nontrivial hexagonal loops and compare it to the experiment in \autoref{fig_NLC} (a). The convergence depends strongly on the parameters. The NLCE data of the specific heat and the error are obtained by the largest two orders:
\begin{align}
    C_{\text{NLCE}}(T) &= C_{\text{NLCE},6}(T)\label{eq:cv_nlce}\\
    \delta C_{\text{NLCE}}(T) &= \text{max}_{T_i > T}\vert C_{\text{NLCE},6}(T_i) - C_{\text{NLCE},5}(T_i)\vert\label{eq:delta_cv_nlce} 
\end{align}

\begin{figure}
    \centering
    \includegraphics{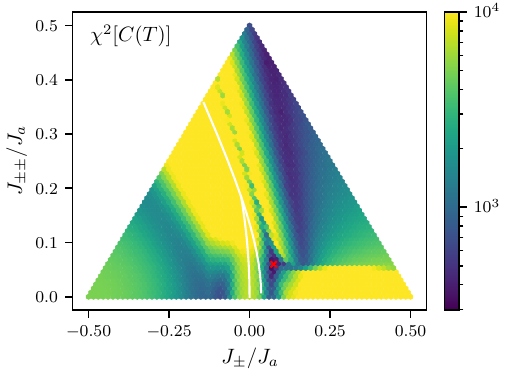}
    \caption{The goodness of fit from comparing the calculated specific heat to the experimental data. The best parameter estimate is marked as a red cross.}
    \label{fig:appendix_cv}
\end{figure}

The error associated with a given parameter set is determined as follows. First, the high temperature tail ($1.9\text{~K} < T < 4\text{~K}$) of the experimental data which exhibits a small error is used to fix the overall energy scale $\alpha$ such that the exchange parameters are scaled to Kelvin. Second, the remaining data points ($T_0 < T < 1.9\text{~K}$) are used to estimate an error for the specific parameter set for different $T_0=0.1,0.125,0.15\text{~K}$. The experiment involves a large systematic error $\delta_{\text{sys}}(T)$ at low temperature due to the uncertainty of the sample mass. To incorporate the systematic error, we introduce a fitting parameter $\varepsilon_c\in[-1,1]$ such that $\sum_{T_i} \left(C_{\text{NLCE}}(T_i) - C_p(T_i)-\varepsilon_c\delta_{\text{sys}}(T_i)\right)^2$ is minimized. The goodness-of-fit for the specific heat data is given by
\begin{align}
    \chi^2[C] = \sum_{T_0<T_i<1.9\text{~K}}\frac{\left(C_{\text{NLCE}}(T_i) - C_p(T_i)-\varepsilon_c\delta_{\text{sys}}(T_i)\right)^2}{{\delta C_{\text{NLCE}}(T_i)}^2+ \delta_{\text{unsys}}(T_i)^2}
\end{align}
where $\delta_{\text{unsys}}$ is the unsystematic error. 
The derived goodness of fit for each parameter set of the exchange constants is shown in \autoref{fig:appendix_cv}.
The best approximation (red cross) defines all three exchange parameters $(J_a,J_b,J_c)$.

\subsection{Magnetic susceptibility} 

Determining the susceptibility in an external field is not as simple as for the specific heat. We generalized the tetrahedral NLCE from Sch\"afer \textit{et al.}~\cite{Schaefer2020} to include the site-dependent Zeeman energies due to the direction of the applied magnetic field. The anisotropy requires calculation of all three independent permutations:
\begin{align}
    (J_x,J_y,J_z) = \left\{ (J_a,J_b,J_c),(J_b,J_a,J_c),(J_b,J_c,J_a)\right\}
\end{align}
In addition to the permutation, we performed the calculation for different values of $\theta\in[0,\pi/2)$. Since we vary $\theta$ which
mixes $S^x$ and $S^z$, we do not need to separately account for parameter permutations which swap $J_{x}$ and $J_z$.

We use the same overall energy scale $\alpha$ extracted from the high temperature tail of the specific heat ($1.9\text{~K} < T < 4\text{~K}$) such that the exchange parameters are in units of kelvin. Then, we apply a field of $\vert \vec{h}\vert=0.01$~T in the (1,1,1) direction such that the easy axis on one site of the tetrahedral unit cell is aligned with the field yielding $\vec{h}\cdot \vec{z_i} =\vert \vec{h}\vert$, while the other sites yield $\vec{h}\cdot \vec{z_i} =-\frac{1}{3}\vert \vec{h}\vert$. We scaled the external field by $k_{\text{B}}^{-1}$ to match the unit of the exchange parameters. We can evaluate the total magnetization using NLCE at finite temperature via
\begin{align}
    M  = g_z\mu_B \sum_i \vec{e} \cdot \hat{\vec{z}}_i
    \left(S_i^{\Tilde{z}}\cos\theta + S_i^{\Tilde{x}}\sin\theta\right),   
\end{align}
\noindent where $\vec{e}$ is aligned with the field with $\vert\vec{e}\vert =1$. The susceptibility is defined by
\begin{align}
    \chi =c\frac{M}{\vert \vec{h}\vert}
\end{align} 
where $c = 9.65$~emu~T$^2$ mol~Ce$^{-1}$ meV$^{-1}$ converts the susceptibility into the correct units.

The susceptibility measurements were performed at higher temperature than the measurements of the specific heat such that fourth order NLCE is sufficient. Since the NLCE data are completely converged for the region of interest, we define the susceptibility and its error the following way:
\begin{align}
    \chi_{\text{NLCE}} = (\chi_{\text{NLCE},4} + \chi_{\text{NLCE},3})/2\\
    \delta\chi_{\text{NLCE}} = (\chi_{\text{NLCE},4} - \chi_{\text{NLCE},3})/2
\end{align}

To begin with we set $g_z=2.57$ in accordance with
expectations for the CEF ground state wave function. 
We include a scaling factor, $\varepsilon_\chi\in[0.5,1.5]$, to allow a variable $g$ factor for the effective system which minimizes the error:
\begin{align}
    \chi^2[{\chi}] = \sum_{T_i}\frac{\left(\varepsilon_\chi\chi_{\text{NLCE}}(T_i) - \chi(T_i)\right)^2}{{\varepsilon_\chi^2 \delta \chi_{\text{NLCE}}(T_i)}^2+ \delta_{\text{unsys}}(T_i)^2}
\end{align}
Note that $\varepsilon_\chi$ scales the (simulated) NLCE data of the susceptibility while $\varepsilon_C$ scales the (experimental) systematic error of the specific heat.
Since $\chi_{\text{NLCE}}\propto g_z^2$, $\varepsilon_\chi$ scales the final result for 
the $g$ factor as $g_z \to \sqrt{\varepsilon_{\chi}} g_z$.

\subsection{Summary of NLCE results} 

We performed extensive NLCE calculations for 2279 parameter combinations ($J_a$, $J_b$, $J_c$) illustrated by the triangle in \autoref{fig:appendix_cv}. In the zero field case, we determined the specific heat up to sixth order which includes nontrivial periodic loops. The region of convergence is comparable to the temperatures in experiments. To account for the systematic error induced by the uncertainty of the sample mass, we added a scaling factor $\varepsilon_c\in[-1,1]$ for each individual calculation. The magnetic susceptibility was calculated for the three permutations of the exchange constants and different angles $\theta\in[0,\pi/2)$. Since the experiments in a magnetic field were performed at higher temperature, a fourth order expansion was sufficient. We additionally added another factor, $\varepsilon_\chi\in[0.5,1.5]$, to include a variation of the $g$ factor.
The best fit result for the $g$ factor was $g_z\approx2.2$, in agreement with Sibille \textit{et al.}~\cite{Sibille2015}.
\begin{figure*}[ht!]
    \centering
    \includegraphics{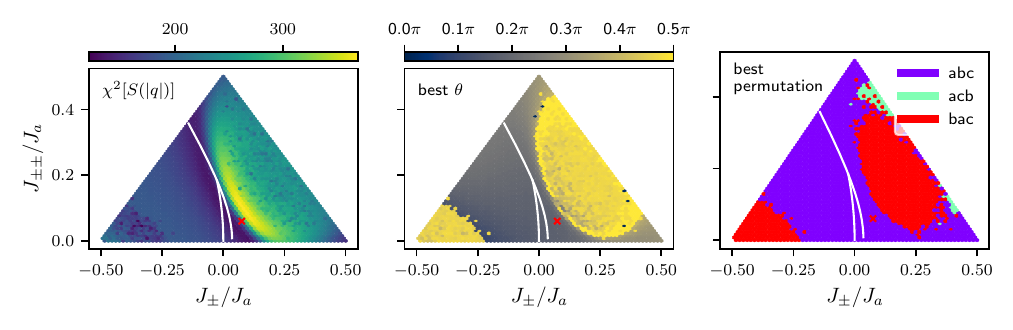}
    \caption{Right: Goodness of fit from comparing the calculated structure factor $S(|q|)$ from powder of \cesno\ to the experimental data. The best fit from fitting both structure factor and specific heat is also indicated. For each pair $(J_{\pm}$, $J_{\pm\pm})$ we show the goodness of fit optimized over both the three permutations of exchange parameters and mixing angles $\theta$. Center (right): value of $\theta$ (exchange permutation) with best goodness of fit for each pair $(J_{\pm}$, $J_{\pm\pm})$.}
    \label{fig:appendix_s_q}
\end{figure*}

\section{Error estimation}

We estimate an error for the best parameter set using a bootstrapping approach. 
For all 2279 parameter sets, we determine the $J_a$, $J_b$, $J_c$, and $\varepsilon_c$ values to agree best with measured heat capacity and $\theta$ and the effective $g$ factor to agree best with the magnetic susceptibility for the respective permutations.

We then generate  $10 000$ samples of two
randomly distributed parameters $r_{\text{NLCE}}\in[-1,1]$ and $r_{\text{exp}}\in[-1,1]$ which we use to vary the NLCE and experimental heat capacity curves within their error bars. For each sample we can then
calculate some new best fit parameters by minimizing:
\begin{align}
    y^2 = \sum_{0.1\text{~K}<T_i<1.9\text{~K}}&\left(C_{\text{NLCE}}(T_i) +r_{\text{NLCE}}\delta C_{\text{NLCE}}(T_i)\right.\\
    &\left. - C_p(T_i)-\varepsilon_c\delta_{\text{syst}}(T_i) - r_{\text{expt}}\delta_{\text{unsyst}}(T_i)\right)^2\nonumber
\end{align}
The experimental data are varied within the unsystematic error $\delta_{\rm unsyst}$, since the systematic mass
error is already accounted for by the parameter $\epsilon_c$.

This procedure then generates a distribution of exchange parameters $J_{\tilde{x}}, J_{\tilde{y}}, J_{\tilde{z}}$. A corresponding distribution of $\theta$ and $g_z$ is then obtained by fitting the susceptibility for each set of $J_{\tilde{x}}, J_{\tilde{y}}, J_{\tilde{z}}$. We do not  consider the errors on the susceptibility calculations and measurements since these are much smaller than the uncertainties on the heat capacity, and so not the dominant source of error in the fits.

The resulting distributions for all optimal values, $J_x$, $J_y$, $J_c$, $\theta$, and $g_z$ are fitted to a normal distribution.
Its standard deviation [permutation $\{a,b,c\}$, cf. \autoref{eq:Js}] defines the error.
\begin{equation}
\begin{aligned}
    (\delta J_{\Tilde{x}},\delta J_{\Tilde{y}},\delta J_{\Tilde{z}}) &= (0.0004, 0.0023, 0.0046)\text{~meV}, \\
    \delta\theta &= 0.0419\pi, \ \  \delta g_z = 0.0067.
\end{aligned}
\label{eq:Js_error}
\end{equation}

\noindent While the above analysis quantifies the uncertainty on the best fit values of the parameters, given the particular model that we use, there is an additional uncertainty coming from the possibility of effects beyond what the model includes (e.g. disorder, further neighbor interactions). This uncertainty is difficult to quantify, but this is likely a bigger source of potential error than the uncertainty in the fitting procedure itself.


\section{Monte Carlo simulations of $S(|\vec{q}|)$}
\label{appendix:MCSim}

We employ classical Monte Carlo simulations to compute the powder-averaged structure factor $S(|\vec{q}|)$.
In contrast to the specific heat, the structure factor does depend on the permutation of exchange parameters, that is on a specific choice of $(J_{\Tilde{x}},J_{\Tilde{y}},J_{\Tilde{z}})$ as well as on the value of the mixing angle $\theta$. Because of this, we can use it to fit the full set of these parameters. To this end, for each of the 2279 pairs of $(J_{\pm},J_{\pm\pm})$, we first run a classical Monte Carlo simulation of the $XYZ$ Hamiltonian \autoref{eq:Ham} at $T = 10$ K and $T = 0.3$ K. We use the energy scale $\alpha$ obtained from an NLCE fit to the high temperature tail of the specific heat to obtain scaled exchange parameters:
\begin{align}
    J_a &= \alpha, \\
    J_b &= 2 \alpha (J_\pm + J_{\pm\pm}), \\
    J_c &= 2 \alpha (J_\pm - J_{\pm\pm}).
\end{align}

In each simulation, we compute the structure factor components
\begin{equation}
    \mathcal S_{ij}^{\alpha\beta}(\vec{q}) = \frac{1}{N}\sum_{\vec{r}, \vec{r'}} \exp[-i\vec{q} \cdot (\vec{r} - \vec{r'})]\expval{s_i^\alpha(\vec{r}) s_j^\beta(\vec{r'})}
\end{equation}
where $s_i^\alpha(\vec{r})$ is the 
$\alpha \in \{a, b, c\}$ component of a classical Heisenberg spin on sublattice $i  \in \{0,1,2,3\}$ in the unit cell at position $\vec{r}$. By a suitable permutation and rotation of components $(J_{a}, J_{b}, J_{c}) \to (J_{\Tilde{x}},J_{\Tilde{y}},J_{\Tilde{z}})$ we can use this quantity to compute the full dipolar-octupolar structure factor $ \mathcal S(\vec{q})$ for all three inequivalent permutations of exchange parameters as well as for 101 values of $\theta$ in $[0, \pi/2]$. Following Sibille \textit{et al.}~\cite{Sibille2019} and Lovesey and Van der Laan~\cite{LOVESEY}, the dipolar-octupolar form factor is included as
\begin{align}
    \mathcal S(\vec{q}) =& \sum_{ij}\exp[-i\vec{q}(\vec{r_i} - \vec{r_j})] \times \nonumber\\
    &\sum_{\alpha, \beta, \gamma, \sigma} 
    \mathcal S_{ij}^{\alpha\beta}(\vec{q}) \mathcal D_i^{\gamma\alpha}(\vec{q}) \mathcal D_j^{\sigma\beta}(\vec{q}) \left( \delta_{\gamma\sigma} - \frac{q_{\gamma}q_{\sigma}}{q^2}\right)
\end{align}
where the form factor $\mathcal D_j^{\alpha\beta}(\vec{q})$ is defined as
\begin{equation}
    \mu^\alpha_j(\vec{q}) = \sum_{\beta} \mathcal D_j^{\alpha\beta}(\vec{q})\,s_j^\beta (\vec{q}),
\end{equation}
where $\vec{\mu_j}(\vec{q})$ is the Fourier transform of the magnetic moment on sublattice $j$. The explicit form is discussed in Appendix F.1. 

We then use the calculated structure factor to compute the differential cross section
\begin{align}
    S(\vec{q}) &= \left(\gamma r_0\right)^2 \times \mathcal S(\vec{q}) \\
    & = \left( 0.290521 \frac{\text{b}}{\text{sr}\,\text{Ce}} \right) \times \mathcal S(\vec{q})
\end{align}
where $\gamma r_0 = 0.539 \times 10^{-12}\,\rm{cm}$ is the magnetic scattering length.

Finally, we average over direction to obtain the powder-averaged structure factor $S(|\vec{q}|)$. Since our simulation is at a finite size, we only sample a discrete set of momenta $\vec{q}$. To perform the powder average, we bin these into momentum shells of width $\mathrm{d}q = 2\pi/(8a)$, where $a$ is the lattice constant, and average over these shells. To compute the goodness of fit, we also rebin the experimental data into the same shells and define
\begin{equation}
    \chi^2\left[ S(|\vec{q}|) \right] = \sum_{q_{\rm min} <|\vec{q}|< q_{\rm max}} \frac{\left(\lambda S_{\rm MC}(|\vec{q}|) - S_{\rm expt~rebinned}(|\vec{q}|)\right)^2}{\delta_{\rm expt~rebinned}^2(|\vec{q}|)},
\end{equation}
where $\lambda$ is an additional scale parameter that is minimized for each dataset as a consistency check and $q_{\rm min} = 0.66\,$\AA$^{-1}$ and $q_{\rm max} = 5\,$\AA$^{-1}$. 
Since we are comparing absolute differential cross sections, $\lambda$ for good fits should be close to one. Indeed we find that for the best parameter estimate [\autoref{eq:Js}], $\lambda = 1.39$. Note that the exact value of this scale factor  $\lambda$ is quite sensitive to the mixing angle $\theta$. To reproduce the exact value of $\lambda = 1.39$ as a best fit, one has to choose the exact best fit value of $\theta = 0.185\pi \approx 0.19\pi$.
We show the goodness of fit, optimized over the three inequivalent permutations of exchange parameters and all values of the mixing angle $\theta$ in \autoref{fig:appendix_s_q}. 
Remarkably, for all good fits, the optimal permutation of exchange parameters is 
$(J_{\Tilde{x}},J_{\Tilde{y}},J_{\Tilde{z}}) = (J_{a},J_{b},J_{c})$, and the optimal value of the mixing angle is $\theta \approx 0.2\pi$. Also the optimal scale $\lambda$ is of order one for all good fits. Hence, while the fit to the structure factor does constrain the exchange permutation and the value of the mixing angle quite reliably, it does not, by itself, strongly constrain the value of exchange parameters.

\subsection{Dipolar-octupolar form factor for 4$f$ electrons}
\label{appendix:DOFF}

The atomic form factor is derived from a multipole expansion of the electric charge density of the 4$f$ electrons,
\begin{align}
    \mathcal D_j = \mathcal Q_{z, j}^{(1)} + \mathcal Q_{x, j}^{(3)} + \mathcal Q_{y, j}^{(3)} + \mathcal Q_{z, j}^{(3)} + \dots.
\end{align}
The $\mathcal Q^{(n)}_{\alpha, j}$ denote the form factor of the $s^\alpha_j$ component of the pseudospin to the form factor at $n$th order in the multipole expansion. 
In particular, $Q_{z,j}^{(1)}$ is the usual dipolar contribution\cite{RauGingrasAnnuRevCMP} which, treating Ce as a hydrogen atom with effective nuclear charge $Z_{\rm eff} = 17$, takes the form
\begin{align}
    \mathcal Q_{x, j}^{(1)} &= \mathcal Q_{y, j}^{(1)} = 0, \\
    \mathcal Q_{z, j}^{(1)} &= \frac{g_z}{2} f_1(\kappa) \hat z_j \otimes \hat e_z.
    \label{eq:form-factor-dip}
\end{align}
Above, $\hat z_j$ is the local easy axis, $\hat e_\alpha$ is a basis vector, and $\otimes$ denotes the outer product between vectors. $\vec \kappa = 2 \vec q a_0 / (a\,Z_{\rm eff})$ with the Bohr radius $a_0$ and lattice constant $a$ ($2.6629$\,\AA~for Ce). Finally, $g_z=18/7\approx 2.57$ is the $g$ factor of the Ce ground state doublet, and the form factor envelope,
\begin{equation}
    f_1(\kappa) = \frac{35 + 364 \kappa^2 - 3760 \kappa^4 + 2880 \kappa^6}{35\,(1 + 4 \kappa^2)^8},
\end{equation}
is related to the integral over the radial charge distribution. 
Note that the dipolar order of the full scattering amplitude as calculated here is not in exact agreement with the dipolar approximation to the form factor \cite{rotter2009formfactor}. For all practical purposes, the difference between \autoref{eq:form-factor-dip} and the dipolar approximation is, however, negligible. This verifies the applicability of the latter in situations where the local moments are dipoles.

The next nonzero order in the multipole expansion, $n=3$, corresponds to the octupole moment of the charge distribution
\begin{align}
    \mathcal Q^{(3)}_{x, j} &= 2c_3\, f_3(\kappa)\, \hat x_j \otimes 
    \mqty(
        24\sqrt 5\, \left[ (\vec \kappa \cdot \hat x_j)^2 - (\vec \kappa \cdot \hat y_j)^2 \right]\\
        -48\sqrt 5\, (\vec \kappa \cdot \hat x_j)(\vec \kappa \cdot \hat y_j)\\
        -168/\sqrt 5\, (\vec \kappa \cdot \hat x_j)(\vec \kappa \cdot \hat z_j)) \\
    \mathcal Q^{(3)}_{y, j} &= 2c_3\, f_3(\kappa)\, \hat y_j \otimes 
    \mqty(
        -48\sqrt 5\, (\vec \kappa \cdot \hat y_j)(\vec \kappa \cdot \hat y_j)\\
        24\sqrt 5\, \left[ (\vec \kappa \cdot \hat y_j)^2 - (\vec \kappa \cdot \hat x_j)^2 \right]\\
        -168/\sqrt 5\, (\vec \kappa \cdot \hat x_j)(\vec \kappa \cdot \hat z_j)) \\
    \mathcal Q^{(3)}_{z, j} &= 2c_3\, f_3(\kappa)\, \left[ \kappa^2 - 3 (\vec \kappa \cdot \hat z_j) \right]\, \hat z_j \otimes \hat e_z
\end{align}
where $c_3 = 3/(14\sqrt{5})$ for Ce is a material-dependent parameter (given on p. 160 of Marshall \textit{et al.}~ \cite{MarshallLovesey1971}), and 
\begin{equation}
    f_3(\kappa) = \frac{9 + 200 \kappa^2 - 240 \kappa^4}{9\,(1 + 4 \kappa^2)^8}
\end{equation}
is related to the integral over the radial charge distribution.

If present, the dipolar contribution to the form factor will dominate at small |{\bf q}|. However, since it is related only to the correlations of $s^{\tilde z}$ and $s^{\tilde x}$ for dipolar-octupolar pyrochlores, this contribution can be suppressed if the correlations of those components are sufficiently weak. In this paper, we hence include both the dipolar ($n=1$) as well as the octupolar ($n=3$) contribution to the neutron scattering cross section.

\section{Quantum Monte Carlo simulations}
\label{appendix:qmc}

\begin{figure}
    \centering
    \includegraphics{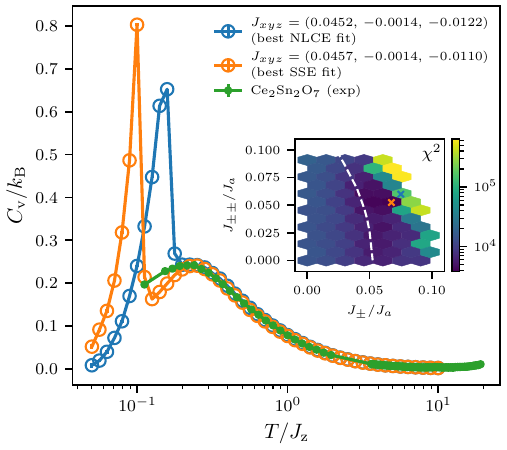}
    \caption{Comparison between specific heat of \cesno\ and as computed for different parameter sets using quantum Monte Carlo simulations. The inset shows the goodness of fit obtained from SSE for parameters around the best high temperature fit. The white dashed line indicates the phase boundary between the ordered and the liquid phase \cite{Huang2020}. Importantly, even when considering all experimentally accessible temperatures the best fits lie in the ordered phase.}
    \label{fig:appendix_sse}
\end{figure}

Since the estimated parameters of \cesno\ [\autoref{eq:Js}] fall into the region in parameter space that allows unbiased quantum Monte Carlo simulations, we implement the stochastic series expansion \cite{sandvik1999} to compute the specific heat for the best parameter fit down to lower temperatures than possible using NLCE. This also allows us to estimate the location of the phase transition that would be expected for these parameters at low temperature \cite{Benton2020, Huang2020}. A quantitative comparison between experimental data and the SSE simulation is shown in \autoref{fig:appendix_sse}. Importantly, the best fit to the high temperature data [\autoref{eq:Js}] does not compare well to the experimental data at low temperatures because the sample would be expected to undergo a phase transition at the lowest measured temperatures already. However, due to the proximity to the phase boundary (indicated by a white dashed line in the inset of \autoref{fig:appendix_sse}), the transition temperature is expected to be very sensititve to the exchange parameters. This is exemplified already by considering the parameters 
\begin{equation}
    (J_{\Tilde{x}},J_{\Tilde{y}},J_{\Tilde{z}}) = (0.0457, -0.0014, -0.0110)\,\text{meV},   
\end{equation}
which differ from the best high temperature fit by only $\sim 10^{-3}$ meV, but this difference suffices to push the critical temperature the lowest experimentally measured temperature. Indeed, performing a SSE parameter sweep in the neighborhood of the best high temperature fit, we find the above parameters to be the best fit when considering the full temperature range of the experimental data. We also stress that even when considering this full temperature range, the best fits all lie within the ordered phase, although very close to the phase boundary to the liquid phase. We show the full goodness of fit $\chi^2_c$ of the specific heat when compared to the SSE simulation result for a limited parameter region in the inset of \autoref{fig:appendix_sse}, where 
\begin{equation}
	\chi_{c}^2 = \sum_{T}\frac{\left(C_{\text{SSE}}(T_i) - C_p(T_i)-\varepsilon_c\delta_{\text{syst}}(T_i)\right)^2}{\delta_{\text{unsyst}}^2}
\end{equation}
and, as for the NLCE fit, $\epsilon_c$ is fitted for each parameter set to minimize $\chi_c^2$.

\bibliography{refs.bib}

\end{document}